# A Multiprocessor Communication Architecture For High Speed Networks

Iyengar. S, N Apte, A.A.Roy, Victoria Jubilee Technical Institute, Bombay, India
S. Sanyal, Computer Systems and Communication Group,
N.M. Singhi, School of Mathematics,
TIFR, Bombay, India
Wu Geng Feng, Dept. of Electronics &Electrical Engg.
USTC, Hefei 230026, Anhui, Peoples Republic of China.

ABSTRACT

Over the years, communication speed of networks has increased from a few Kbps to several Mbps, as also the bandwidth demand, Communication Protocols, however have not improved to that extent. With the advent of Wavelength Division Multiplexing (WDM), it is now possible to "tune" protocols to current and future demands. The purpose of this paper is to evolve a High Speed Network architecture, which will cater to the needs of bandwidth-consuming applications, such as voice, video and high definition image transmission.

INTRODUCTION

Communication speeds of local-, metropolitan-, and wide- area networks have increased by about six orders of magnitude[1]. Current technology and WDM networks[2,3] offer bandwidths higher than peak electronic data-rates. At the same time, user-demand for bandwidth has increased. Given the fast-increasing utilization of networks for high definition image transmission, digital voice and video, future networks have to meet demands of over 1Gb/user[2].

Even though current media can sustain such a demand, existing protocol structures and implementations are still unable to exploit media to cater these needs. The discrepancy between the speed of communication processor and the channel, translates into bottlenecks in the communication process. We suggest an approach to exploit high speed media to deliver a high throughput using multiprocessors for faster packet-processing.

THE SWIFT ARCHITECTURE[4]

1. Description

The SWIFT architecture as detailed by Chlamtac I. and Ganz A. forms the backbone of our design. The architecture works at three levels. At the physical layer it uses a multichannel physical configuration. It provides b, $1 <= b <= N$, (N is total number of nodes), sub-channels[5] of equal bandwidth. The data link layer protocol is slot-oriented with exactly b permissions in each slot, equal to number of sub-channels. Sub-channel allocation depends upon node address and a slot-allocation matrix. Multichannel network layer is responsible for routing packets and delay factors influence routing. The FINE algorithm deals with the delay calculations. The architecture favours broadcast networks.

2. Drawbacks

Though the architecture potentially improves system capacity compared to conventional single-channel architectures, it can be improved further with optical fibre technology and multiprocessors. SWIFT protocols do not permit full-duplex (FDX) communication since each node can transmit only to a particular node during its slot-time; it can not receive from the same node (Fig.1). Hence, they can only be of stop-and-wait type. Window or similar protocols used in most practical networks[6] can not be mapped easily on SWIFT. Such an implementation would mean longer time-out periods and hence reduced performance. Thus SWIFT manages to utilize more bandwidth out of available bandwidth than other architectures, but it still leaves room for improvement in terms of FDX communication between nodes.

In this paper, we therefore propose an architecture for high speed networking which can sustain various traffic requirements. It essentially employs a fibre optic network and uses multiprocessors to remove the aforementioned bottlenecks. The architecture, termed, Multiprocessors In Networking (MIN), manages bandwidth-consuming applications using demand-oriented channel-allocation.

THE MIN ARCHITECTURE

1. Channel Allocation Policy

The main aim of MIN is to facilitate FDX communication for every node. In FDX communication we must allocate a pair of sub-channels between the two nodes. The connection requirements using WDM are as shown in Fig.2.

We note that for an N-node network, we require N sub-channels. These sub-channels are sorted in terms of time; in each time slot a node can communicate with only one other node which is distinct from the nodes it can communicate with, in the other slots. Thus the requirements (FDX and TDM-WDM) for allocation matrix A are:

A: Destination-conflict freeness.
B: Single node-pair allocation per pair of sub channels.
C: Non-overlapping pairing in each slot.
D: Unique destination (from previous slots) in each slot.
E: Equi-spaced allocation (fixed cycle time) per node pair.

Cycle time is defined as number of elapsed slots between two slots with permissions for the same set of node-pairs. Any construction of A will favour an even number of nodes.

*Construction of the sub channel allocation matrix A:*
We note that A will have N rows (one row per node) and (N-1) columns, since $^NP_2$ distinct pairs are possible; the cycle time for each node will be (N-1) slots. We number the nodes from 0 through (N-1) and the equal-duration slots from 1 through (N-1). We have, for elements of A:
  $a_{kj}=i$ iff $a_{ij}=k$ , $0<=i, k<=(N-1); 1<=j<=(N-1)$ i.e., FDX communication is possible between the nodes i and k in the slot j. The matrix A can be constructed from another (N-1)x (N-1) matrix B where B is a Latin square[7] B can be constructed as:
  $b_{ij}=(i+1)\bmod(N-1)$
Matrix A can be constructed using the mappings:
If $b_{ik}=j$          $i!=j, j!=2i, i!=(N-1)!=j$
Then $a_{ij}= k$       $0< i < N-1, 1<=j<=N$
$a_{ij}=0$             $j=2i, 0<i<=N-1, 1<=i<=N$
$a_{0j}=i'$            where $a_{i'j}=0$
$a_{ij}=N-1$           $0<i<N-1, j=i$
$a_{N-1,N-1}=0$        $i=j=N-1$
The matrix A for an 8-node network is shown in Fig.3.
For an odd number of nodes matrix A' of N'X N', where N' is odd can be constructed as follows:
Choose N=N'+1 (N is even). Construct N X (N-1) matrix as above. Drop row number N' and replace all entries in A equal to N' by 'X' indicating idle state for that companionless node.
The cycle-time for N-node network can be given as:
  $T=2*[N/2]-1$……………….(I)
This allocation follows FDX communication in every slot.

2. Routing
In MIN, we have three different levels of routing; with increasing complexity in operation. The concept of routing is based on the idea that a packet from i to k may be routed via j if the slots i-j and j-k occur before the slot i-k.

To formally describe the algorithm, we define waiting-time delay $d_w$ for a node to be the number of slots from the current slot – when a packet for a particular destination was received – to the slot in which the transmission actually occurs.
  $d_w=N-1-|t - t'|$………………..(II)
where, t is current time slot and t' is the time slot when transmission actually occurred.
Lemma 1: The maximum waiting-time delay ( in terms of slots ) experienced by a node for transmission to a particular destination is N-2.
Proof: From (II) we see that
  $d_{wmax}=(N-1)-|t – t'|_{min}$
Such a condition occurs when a node just misses its designated slot and no other intermediate node is available for forwarding its packets. Hence,
$| t – t' |_{min}=1$
which gives $d_{wmax}=N-1 - 1=N-2$ ……………….(III)

*Level 0 Routing;* This level just follows the channel allocation matrix A for transmission. It does not allow any other transmission permission. As in (III), maximum delay experienced will be N-2. This type of routing may result in buffer overflow during large file transfers.

*Level 1 Routing;* Here we use other nodes for routing packets. Intermediate courier nodes are chosen according to a routing table RT , which can be constructed as follows:
          {1 if delay is reduced when i transmits packet
          {for j, in slot t , to current node
  $rt_{ijt}$ = {
          {0 i waits for an optimal slot or it uses
          {level 0 or level 2 routing

An algorithm for constructing matrix RT ( N X N X (N-1)) is given in Fig.4.

Initialize matrix RT to 0
   FOR each source node i ( 0<=i<=N-1 )
   FOR each destination node d ( 0<=d<=N-1, d!=i )
    FOR each slot t i ( 1<=t<=N-1 )
     BEGIN
     dw = current waiting-time delay
     IF dw = 0 THEN exit
     ELSE BEGIN
    dmin = dw
    trans = 0
    END
    REPEAT
     Probe packet walk-path for source node in allocation matrix to get next intermediate node. Replace source node with the intermediate node. For every replacement, trans=trans+1
     UNTIL destination is reached
    IF trans < dwmin THEN BEGIN
     dwmin = trans
     t' = t
     END
  END
IF dwmin < N-2 THEN   $rt_{ijt'}$ = 1

**Fig. 4 Level 1 Routing**

Thus if a source node misses its designated slot for a particular destination node it scans the time-axis of the corresponding entry in RT for an optimal slot. If there is none, then it uses Level 0 or Level 2 Routing.

*Level 2 Routing;* Here we require additional N sub-channels for proper synchronization of events and an extra set of transmitters and receivers at every node.
Consider slot A-B. In case both A and B do not want to communicate with each other, this slot will be empty. We can make use of this slot. In the level 2 routing, A disconnects its transmit sub-channel with B and uses it to connect to C which is some other node A wishes to communicate with; A has missed the designated slot A-C. A will first convey to C its intention to transmit (Fig.5). The decision about continuing the session depends upon C, which will base its decision on the number of requests it has received. Requests for sessions from various nodes will be honored depending on the priority of the node which must be predefined (e.g. priority based on node addresses). Thus C will communicate its decision to the proper node which then is eligible to transmit in this slot, to C. The connection between the node-pairs about their respective destination (i.e. in case both A and B wish to communicate with C) can be resolved based on a similar priority scheme. To minimize this overhead, node-pairs use status packets at start of the slot to decide the course of action. The resulting structure of the slot is shown in Fig.6.

This type of interaction however must be supported by the hardware – we need to have receivers (at least) for implementation of this routing. The processing power of the network interface will put an upper limit on the number of receivers, as the node now have to manage multiple sessions. All receivers used a fixed tone but transmitters are tunable, to allow transmission to different fixed receivers. To manage these protocols we have to use multiprocessors.

3. The role of multiprocessors

It was desired that system performance be comparable to the bandwidth of physical media. As speed and bandwidth of the media increase it is the software that influences overall packet delay. Efficient functioning of the system will be achieved only if hardware and software are properly matched.

An important observation is that protocols based on extensively layered architecture are inherently inefficient for high speed communication[8]. Layered approach works satisfactory in normal cases; but, a strict adherence to it increases processing and hence queuing delays. Reasons for this degradation include replication of functions in various layers, overhead of control messages, evaluation of functions which are not necessary in all situations etc. Multiprocessors can be used[9] if packet-processing is parallelized. For networks with fibre links and digital switches the error rate is low. So success-oriented protocols with forward error correction[10] can be used, thus minimizing the delay in most of the cases but making the recovery delayed in few others. This necessitates restructuring of OSIRM[1,8].

With respect to packet-processing the relation between various processes can be identified as follows[1,8].
A: Parallel – Processes which can run independent of each other e.g. flow control and decryption.
B: Concurrent – Processes which work on same data. Semaphores control access to various sections of data.
C: Pipelined – The set of functions which act upon output from some other functions in the set.

This taxonomy helps us to use multiprocessors. We note that send and receive sections of a node can function in parallel. They have limited interaction between themselves, such as during a time-out, when receive-section intimates transmit-section about not receiving an acknowledgement for a packet sent. But such occasions are rare with success-oriented protocols. The multiprocessor architecture that emerges for the transmit section is shown in Fig.7.
We use identical processors for various functions. We first define the relationship between the functions. We assume that $fn_{11}$ and $fn_{21}$ are pipe-linable functions while $fn_{11}$ and $fn_{12}$ can run in parallel. These functions use a shared memory buffer to store processed packets. A communication controller then uses this buffer to take appropriate packets and puts them in a queue. A slot-filter then chooses packets eligible for transmission during the current slot. A routing-filter then acts upon these packets for proper routing based on the routing policy.

CONCLUSION

In this paper, we have evolved an architecture which would cater to the needs of current and future bandwidth-consuming applications in high-speed networking. The changes in design of network interfaces are expected to improve the system throughput and average packet delay.

Current concepts and implementations of protocols being inefficient for future networks, we have tried to eliminate the bottlenecks at various levels. These changes have strongly been influenced by WDM and related technology.

We have proposed a TDM-WDM channel allocation policy which can serve multiple users. Such an allocation will lead to optimal utilization of channel. The policy also facilitates FDX communication which will further help in improving performance. As this simple allocation will cause most of the bandwidth going wasted we have developed two more levels of routing. Such a routing mechanism can be made sensitive to the traffic and can adapt its behavior accordingly. At network interface level, a multiprocessor architecture emerges as the only practical solution.

We believe that the ideas presented in the paper will initiate a discussion in future networking, looking from this aspect.

```
    1  2  3  4  5  6  7
 1  2  3  4  5  6  7  8
 1  2  3  4  5  6  7  8
 1  2  3  4  5  6  7  8
 1  2  3  4  5  6  7  8
 1  2  3  4  5  6  7  8
 1  2  3  4  5  6  7  8
```

Fig 1: Subchannel Allocation matrix of SWFT

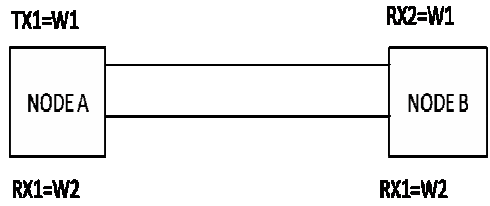

Fig 2: FDX Communication using WDM

```
    1  2  3  4  5  6  7
 0  4  1  5  2  6  3  7
 1  7  0  2  3  4  5  6
 2  6  7  1  0  3  4  5
 3  5  6  7  1  2  0  4
 4  0  5  6  7  1  2  3
 5  3  4  0  6  7  1  2
 6  2  3  4  5  0  7  1
 7  1  2  3  4  5  6  0
```

Fig 3: Subchannel Allocation Matrix of MIN

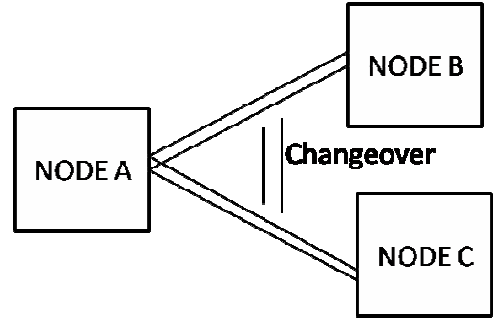

Fig 5 Level 3 Routing

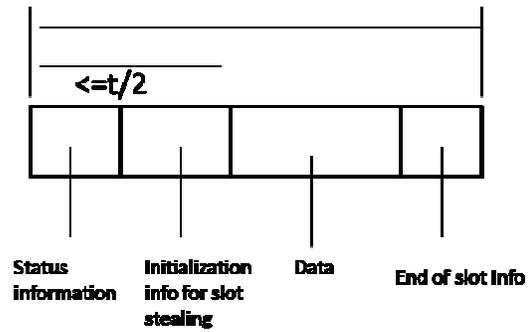

Fig 6: Structure of Slot

Fig 7: Architecture for communication subnet

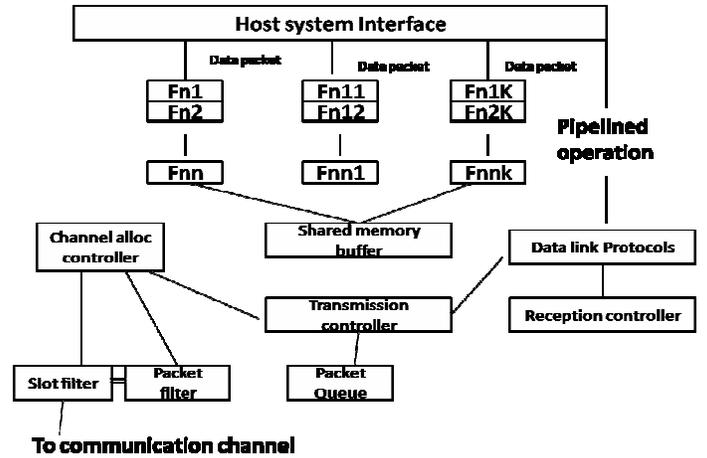